\documentclass[final,5p,times]{elsarticle}
\usepackage{amssymb}
\usepackage{amsmath}
\usepackage{lineno}
\usepackage{graphicx}
\RequirePackage{xspace}
\usepackage{hyperref}

\def\epem {\ensuremath{e^+e^-}\xspace}
\def\emem {\ensuremath{e^-e^-}\xspace}

\newcommand{\kev}{\ensuremath{\mathrm{\,ke\kern -0.1em V}}\xspace}
\newcommand{\mev}{\ensuremath{\mathrm{\,Me\kern -0.1em V}}\xspace}
\newcommand{\mevcc}{\ensuremath{{\mathrm{\,Me\kern -0.1em V\!/}c^2}}\xspace}
\newcommand{\gev}{\ensuremath{\mathrm{\,Ge\kern -0.1em V}}\xspace}
\newcommand{\gevcc}{\ensuremath{{\mathrm{\,Ge\kern -0.1em V\!/}c^2}}\xspace}
\newcommand{\tev}{\ensuremath{\mathrm{\,Te\kern -0.1em V}}\xspace}
\newcommand{\tevcc}{\ensuremath{{\mathrm{\,Te\kern -0.1em V\!/}c^2}}\xspace}
\newcommand{\ev}{\ensuremath{\mathrm{\,e\kern -0.1em V}}\xspace}

\def\mum  {\ensuremath{{\,\mu\rm m}}\xspace}

\def\cms  {\ensuremath{{\rm \,cm}^{-2} {\rm s}^{-1}}\xspace}

\newcommand{\E}{\ensuremath{\varepsilon}\xspace}

\newcommand{\be}{\begin{equation}}
\newcommand{\ee}{\end{equation}}
\newcommand{\bc}{\begin{center}}
\newcommand{\ec}{\end{center}}
\newcommand{\bi}{\begin{itemize}}
\newcommand{\ei}{\end{itemize}}
\newcommand{\ben}{\begin{enumerate}}
\newcommand{\een}{\end{enumerate}}

\journal{Nuclear Instruments and Methods in Physics Research A}

\begin{document}

\begin{frontmatter}



\title{Energy recovery twin linear $e^+e^-$, $e^-e^-$ colliders (ERLC ) with high luminosities and accelerating gradients}


\author[label1,label2]{V.I.~Telnov} 

\affiliation[label1]{organization={Institute of Nuclear Physics},
            addressline={prosp. Lavrenieva 11},
            city={Novosibirsk},
            postcode={630090},
            country={Russia}}
\affiliation[label2]{organization={Novosibirsk State University},
            addressline={Pirogova str. 2},
            city={Novosibirsk},
            postcode={630090},
            country={Russia}}
\ead{telnov@inp.nsk.su}

\begin{abstract}

A recently proposed superconducting linear collider with energy recovery (ERLC) and multiple beam reuse employs twin RF structures to eliminate parasitic collisions in the linacs. Such a collider can operate in either pulsed or continuous-wave (CW) mode, achieving a luminosity of
 ${\cal O}(10^{36})$ cm$^{-2}$s$^{-1}$ at $2E_0$\,=\,250--500 GeV. This paper demonstrates that in pulsed mode, the ERLC luminosity is independent of the accelerating gradient for a fixed total power, enabling operation at the highest available gradients. A similar independence holds for the CW mode when the available power significantly exceeds the operational threshold. The luminosity scales with the cavity quality factor as $L\propto Q_0^{1/2}$. We also present, for the first time, a study of a twin $e^-e^-$ ERLC and estimate its performance. This configuration is simpler than the $e^+e^-$ version as it eliminates the need for beam recirculation; electrons can be generated anew for each cycle. In this case, the luminosity scales as $L\propto Q_0^{1/4}$. Furthermore, the use of traveling-wave (TW) RF structures allows for higher gradients and reduced thermal loading. We show that an ERLC with $G$\,=\,40 MeV/m can operate in CW mode, reaching luminosities of $L_{e^+e^-}$=\,(1--2.5)$\times 10^{36}$ and $L_{e^-e^-}$=\,(3--7)$\times 10^{36}$ cm$^{-2}$s$^{-1}$ at $2E_0$\,=\,250 and 500 GeV, respectively, with a total power consumption of 150--300 MW. These results position the ERLC as a highly promising candidate for a future Higgs factory.

\end{abstract}


\begin{keyword}
superconducting linear collider \sep energy recovery  \sep $e^+e^-$ colliders \sep $e^-e^-$ colliders \sep  accelerating gradient \sep twin RF structures \sep Higgs factory




\end{keyword}

\end{frontmatter}

\section{Introduction}

As early as 1965, M. Tigner~\cite{Tigner} proposed superconducting linear colliders as an alternative to the circular
 machines then planned for energies of 3--4 GeV. He noted that in a superconducting collider, after passing the interaction point (IP), the beams could be decelerated in the opposing linac, returning their energy to the accelerating structures. In the 1970s, linear colliders gained attention as a means to reach even higher energies, since energy losses from synchrotron radiation in circular colliders scale as $E^4/R$ per turn. In 1975, U.~Amaldi published the first paper on high-energy linear colliders~\cite{Amaldi}, independently proposing a superconducting design with energy recovery. However, by the late 1970s, it became evident that this approach faced a fundamental limitation: accelerating and decelerating beams would collide within the linacs, leading to beam disruption. Avoiding these parasitic collisions by having only one beam in the linac at a time resulted in a collision rate, and thus a luminosity, that was too low~\cite{Gerke}. Consequently, research shifted toward single-pass linear colliders, which could achieve significantly higher luminosity. All subsequent linear collider projects have followed this single-pass paradigm.

In the 1990s, several linear collider (LC) projects were under development, including VLEPP, NLC, JLC, CLIC, TESLA, etc.~\cite{Loew}. Since 2004, the field has converged on two primary designs: the ILC~\cite{ILCTDR,ILC} and CLIC~\cite{CLIC}. The ILC utilizes superconducting (SC) niobium technology, following the TESLA design, whereas CLIC employs copper cavities operating at room temperature. Both projects aim for a luminosity of approximately 2$\times 10^{34}$\cms at a center-of-mass energy of $2E_0$\,=\,250 GeV, which is optimal for Higgs boson studies.
Although the ILC has been ready for construction for many years, a final decision on the next collider to follow the LHC is still pending. In the meantime, new strategies have emerged. For instance, circular Higgs factories such as FCC-ee~\cite{FCC} and CEPC~\cite{CEPC} propose luminosities that are an order of magnitude higher at these energies. Moreover, \cite{Litvinenko} proposed an \epem collider with energy recovery in a 100-km ring tunnel, claiming a luminosity several times higher than that of the FCC-ee. However, these estimates were found to be incorrect. A comprehensive review of these competing approaches was prepared for the Snowmass-2021 conference~\cite{Faus}.

\begin{figure*}[!htb]
\centering
\includegraphics[width=14cm]{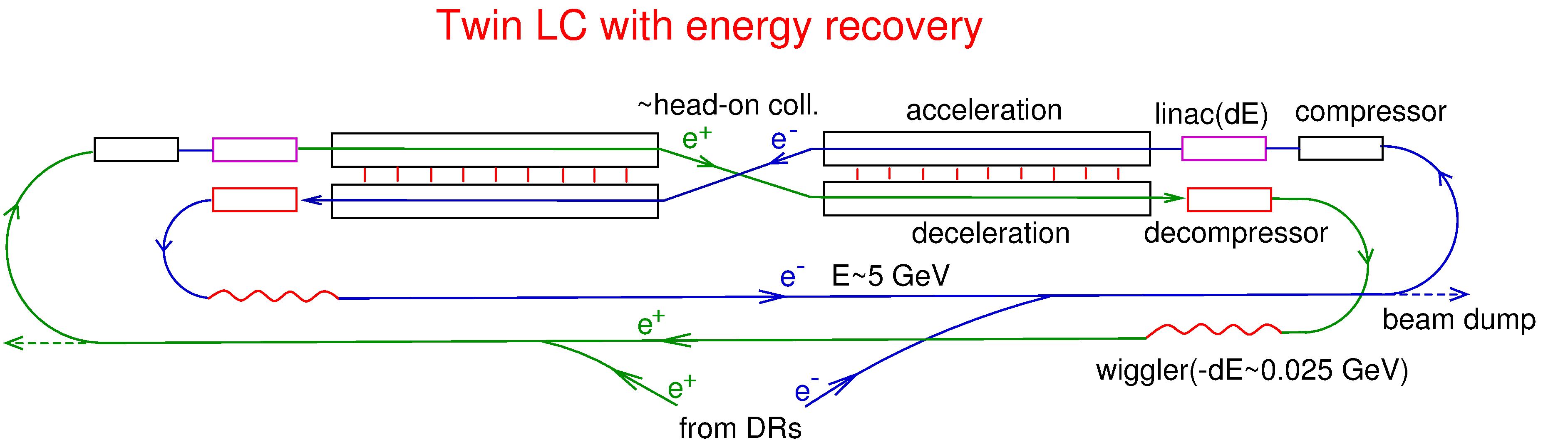}
\caption{The layout of the SC twin \epem linear collider~\cite{Telnov-erlc}}.
\label{twin-scheme}
\vspace{-4mm}
\end{figure*}

 In this paper, we further develop my recent proposal for a \underline{twin} superconducting \epem linear collider with energy recovery (ERLC)~\cite{Telnov-erlc}, where the problem of parasitic collisions in linacs is eliminated. Its layout is shown in Fig.~\ref{twin-scheme}. In this \epem collider, beams are accelerated to $2E_0$\,=\,250--500 GeV and then decelerated to \mbox{$E \approx 5$} GeV in separate, parallel linacs with coupled RF systems.  In this configuration, accelerating and decelerating beams do not collide, allowing bunches to follow one another at very short intervals, down to the RF wavelength. RF power is supplied to the beams both from external sources and from the decelerating beam. The  e$^{+}$ and e$^{-}$ are reused many times (>$10^5$), fully exploiting the advantages of superconductivity. Low-emittance beams are prepared in damping rings and injected into the collider at an energy of approximately 5 GeV. Energy spread induced during collisions is suppressed by wigglers installed in the return path (at 5 GeV). The relative energy loss in wigglers is about 0.5\%, and the steady-state equilibrium energy spread at the IP due to beamstrahlung is below 0.2\% -- comparable to ILC and CLIC parameters before collision. Such a collider operates similarly to a cyclic machine with a low disruption parameter; the emittance growth during collisions is minimal and compensated by the wigglers. As the beam decelerates to 5 GeV, its relative energy spread increases. To ensure loss-free transport through the arcs, this spread is reduced by a factor of 10--15 using a bunch (de)compressor. The achievable \epem luminosity, ${\cal O}(10^{36})$ \cms, significantly exceeds that of the single-pass ILC and is several times higher than at FCC or CEPC.

 In our previous article~\cite{Telnov-erlc} a relatively low accelerating gradient of $G$\,=\,20 MeV/m was assumed. This might be perceived as a disadvantage, given that the ILC aims for approximately 35 MeV/m. However, it has recently been noted that gradients can be further increased by employing traveling wave (TW) structures instead of standing wave (SW) ones~\cite{Belomestnykh}. This approach could potentially enable gradients of 70 (Nb) to 100 (Nb$_3$Sn) MeV/m, though the technology is still in development.

\begin{figure}[!htb]
\centering
\includegraphics[width=8.5cm]{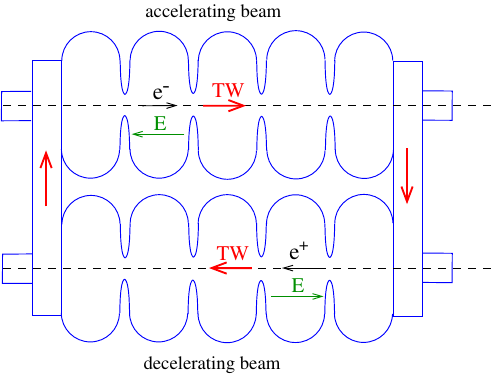}
\caption{Twin ERLC with coupled parallel traveling-wave (TW) accelerating cavities. The upper cavity accelerates the electron beam  while the lower cavity decelerates the oppositely moving positron beam.}
\label{cavity-1}
\end{figure}

\begin{figure}[!htb]
\centering
\includegraphics[width=5.5cm]{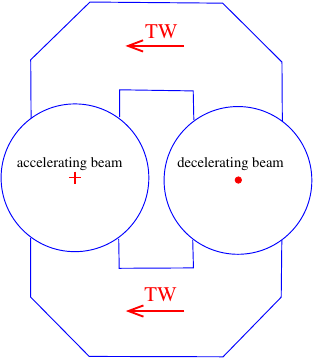}
\caption{Coupling of traveling-wave (TW) cavities with symmetric waveguide pair to eliminate the transverse fields at the beamlines.}
\label{cavity-2}
\end{figure}

   The ERLC design requires a very rapid energy flow between the parallel linacs; this process is inherently facilitated by traveling wave acceleration. Consequently, this scheme is naturally suited for achieving such high accelerating gradients~\cite{Belomestnykh,Yokoya}. A possible configuration of twin ERLC TW cavities is shown in Fig.~\ref{cavity-1}. However, coupling the linacs with a single waveguide introduces an asymmetry in the RF field, creating a transverse force on the $e^+/e^-$ beams. To eliminate this effect, energy can be transferred through symmetric waveguide pairs directed upward and downward, as illustrated in Fig.~\ref{cavity-2}.


    In this paper, we demonstrate that the ERLC collider is also compatible with high accelerating gradients. In pulsed mode, the luminosity is independent of the gradient for a fixed total power, varying only slightly due to the dependence of the cavity quality factor $Q_0$
 on the gradient $G$. We also present, for the first time, a study of the \emem ERLC configuration. Such an \emem collider is significantly simpler than the \epem version, as it does not require beam recirculation and can achieve higher luminosity. Finally, we show that both \epem and \emem
 ERLC colliders can operate in continuous-wave (CW) mode at $2E_0$\,=\,250--500 GeV, reaching luminosities above $10^{36}$\cms by utilizing Nb$_3$Sn and TW RF structures.

\section{ ERLC-\epem: dependence of $L$ on $G$ and $Q_0$}

  In this section, we follow the approach described in Ref.~\cite{Telnov-erlc}, where the necessary formulas were derived; however, the dependence on the accelerating gradient was not explicitly emphasized.

       The power consumption is dominated by two main contributors~\cite{Telnov-erlc}:
        \bi
       \item RF losses in the superconducting (SC) cavities: This po\-wer is required to cool the low-temperature cavities and is independent of the number of particles per bunch. The energy dissipation in a single multi-cell cavity is given by $P_{\rm RF,dis}=V_{\rm acc}^2/(R/Q)/Q_0$, where $V_{\rm acc}$ is the operating voltage. The R/Q ratio (shunt resistance divided by the quality factor) depends solely on the cavity geometry. The cavity quality factor $Q_0$ (exceeding $10^{10}$ in current LC projects) is determined by the residual surface conductivity of the SC material, which in turn depends on the RF frequency.
           The overall refrigeration efficiency for heat transfer from the SC temperature $T_{\rm SC}$ to room temperature $T_{\rm r}$ is $\eta=\varepsilon T_{\rm SC}/(T_{\rm R}-T_{\rm SC})$, where $\varepsilon$ is the technical efficiency, typically reaching 0.1--0.3 for $T_{\rm SC}$\,=\,1.8--4.5 K. \\[-6mm]
        \item  High-Order Mode (HOM) losses: This power is required to compensate for and remove HOM losses. The HOM energy loss of a bunch per unit length is approximately $2e^2N^2/r_a^2$, where $N$ is the number of particles per bunch and $r_a$ is the cavity aperture radius. For a bunch spacing $d$, the HOM power scales as $P_{\rm HOM} \propto N^2/d$. Most of this energy can be extracted from the SC cavities in two ways: a) using HOM couplers that dissipate energy at room temperature, or b) using specialized HOM absorbers located between cavities and maintained at an intermediate temperature (around 80 K), where refrigeration efficiency is high. However, a small fraction of the HOM energy is inevitably dissipated in the SC cavity walls at 1.8--4.5 K.\\[-6mm]

        \item To highlight the primary scaling laws, we neglect less significant power consumers, such as beam generation, radiation in damping wigglers and from beamstrahlung, and static cryogenic losses.
              \ei

First, consider the case where the collider operates in pulsed mode with a duty factor DF. This mode is feasible at any average power level and at any high accelerating gradient. The total power (considering only the main contributions) is given by:
    \be
    P_{tot}= \left(k_1 + k_2 \frac{N^2}{d}\right) \times DF,
    \label{abd}
    \ee
where the coefficients $k_1$ and $k_2$ (corresponding to'$a$' and '$b$' in Ref.\cite{Telnov-erlc}) describe the RF and HOM losses, respectively. Both coefficients are proportional to the length of the collider (and thus to $E_0$).

The \epem luminosity per bunch collision $L_1$ is determined by the collision effects (beam radiation and bunch instability). Let us consider the collision of two flat beams where the luminosity is limited by these effects. If we were to place two pairs of such colliding beams side-by-side in the horizontal plane, the collision effects would remain unchanged because the field of a flat beam is proportional to $N/\sigma_x$, which stays constant in this configuration. In this case, the total luminosity would double, as would the total number of particles in the combined beam; hence, $L_1\propto N$.

The total luminosity is given by:
\be
L \propto \frac{N}{d} DF =\frac{N}{d}\left( \frac{P}{k_1+ k_2N^2/d}\right).
\label{lume+e-1}
\ee
The luminosity reaches its maximum,
\be
L\propto \frac{P}{2\sqrt{k_1k_2 d}},
\label{lumdc}
\ee
under the conditions:
\be
N=\sqrt{\frac{k_1d}{k_2}}\quad \text{and} \quad DF= \frac{P_{\text{tot}}}{2k_1}.
\label{lumdc-1}
\ee
Thus, the luminosity is maximized when the power consumption for RF losses and HOM losses are equal.

From Eqs.(\ref{lumdc}) and (\ref{lumdc-1}), we observe that for a fixed total power:
\ben
\item $L\propto 1/\sqrt{d}$,  implying that the bunch spacing should be as small as possible (ideally $d=\lambda_{\rm RF}$).

 \item $L\propto \sqrt{Q_0}$, since the RF loss coefficient scales as $k_1 \propto 1/Q_0$.

  \item The luminosity $L$ is independent of the accelerating gradient $G$.
 This follows from the fact that the collider length scales as $l_c \propto 1/G$. Consequently, the HOM loss coefficient scales as $k_2 \propto l_c \propto 1/G$, while the total RF losses scale $k_1 \propto G^2 \times l_c \propto G$. Substituting these into Eq.(\ref{lumdc}) yields a constant $L$.
 \item The optimal parameters scale as $N$\,=\,$\sqrt{k_1d/k_2}$\,$\propto$\,$G\sqrt{d/Q_0}$, $DF$\,$\propto$\,$1/G$.
 \een
 These results demonstrate that the path toward high accelerating gradients in the ERLC is now open. The luminosity remains independent of $G$ and shows only a weak dependence on $Q_0$ -- an unexpected result arising from the optimized scaling of $N$ and $DF$.
  At a gradient of $G$\,=\,20 MeV/m, the optimal bunch population is approximately \mbox{$N \sim 10^9$}~\cite{Telnov-erlc}. Consequently, $N$ can be readily increased by several times to maintain optimality at even higher accelerating gradients.

   In continuous-wave (CW) mode, the power and luminosity relations are:
 \be
 P= k_1 + k_2 \frac{N^2}{d}, \;\;\;\; L \propto \frac{N}{d},
 \ee
  which yields:
\be
N=\sqrt{\frac{(P-k_1)d}{k_2}}, \;\;\;\;\;\; L\propto \sqrt{\frac{(P-k_1)}{k_2 d}}.
\label{lumcw}
\ee
Continuous-wave (CW) operation requires a threshold power $P>k_1$, where $k_1 \propto G$. However, at power levels significantly exceeding this threshold, the behavior becomes similar to that of the pulsed mode: the specific luminosity $L/P$ is independent of $G$ and remains proportional to $Q_0^{1/2}$.

\begin{figure*}[!ht]
\centering
\includegraphics[width=14cm]{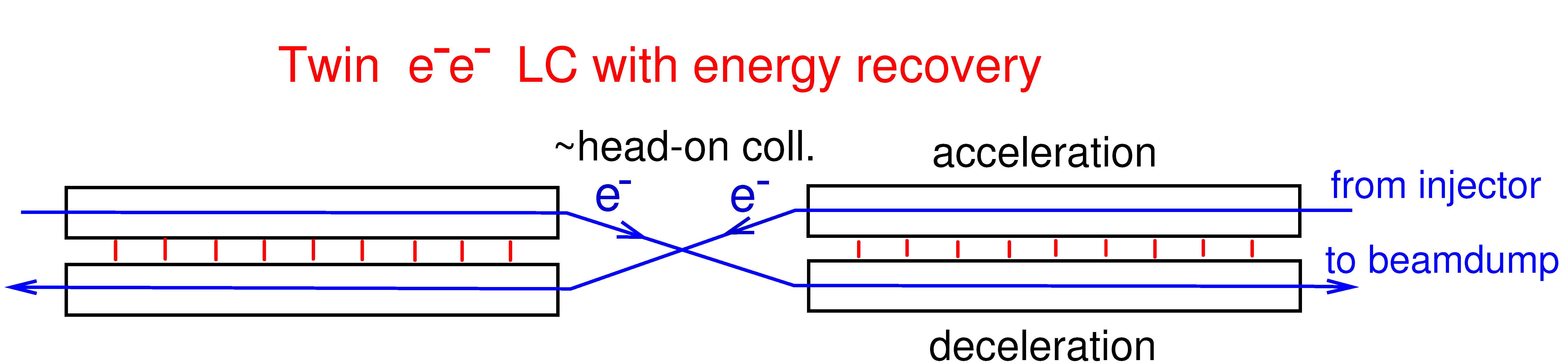}
\caption{The layout of the SC twin \emem linear collider.}
\label{ee-scheme}
\end{figure*}

 \section{ERLC-\emem}

 Next, let us consider a twin \emem linear collider with energy recovery. Such a machine is of great scientific interest; its physics program has been explored in several dedicated workshops, and this configuration has consistently been an integral part of the discussion for future linear collider projects. A sche\-matic layout of the \emem ERLC is shown in Fig.~\ref{ee-scheme}.

This configuration is significantly simpler than an \epem collider because it eliminates the need for beam recirculation. Low-emittance electron beams can be generated anew for each cycle, bypassing the need for damping rings. Furthermore, since the beams are used only once, they can be more tightly focused and allowed to undergo higher disruption than in the \epem case.

The key difference between the ILC and the ERLC is that in the latter, beams return their energy to the RF field after collision. However, energy recovery is not total; a portion of the energy is lost to beamstrahlung at the interaction point (IP) and must be replenished by the RF system. (In an \epem ERLC with multiple beam reuse, such average energy losses are typically negligible).

In addition to the average power loss, collisions induce a broad energy spread with long tails. To effectively recover the beam energy, a decompressor must be installed at the end of the linac to narrow this spread. The lowest-energy particles can then be removed using a series of mini-beam dumps. For simplicity, we assume that twice the average energy loss must be compensated, with an RF power efficiency of \mbox{$\eta_{\text{rf}} \sim 50$} \%.

 The relative energy loss due to beamstrahlung is given by \cite{Chen-Yokoya}:
 \be
\delta=\frac{\Delta E}{E_0} \approx \frac{5r_e^3N^2\gamma}{6\sigma_z\sigma_x^2},
\ee
 where $r_e=e^2/mc^2$ is the classical electron radius, $\gamma =E_0/mc^2$, and $\sigma_z$, $\sigma_x$ are the bunch length and the horizontal beam size at the IP, respectively.

 The power required to compensate for these energy losses in both beams, including the factor of two for the safety margin mentioned above, is given by
 \be
P_{rad}= \frac{4NcE_0 \delta}{\eta_{\text{rf}}d}=k_3\frac{N^3}{\sigma_z\sigma_x^2d}, \,\,\,\,\,\,\, k_3=\frac{10 E_0r_e^3\gamma c}{3\eta_{\text{rf}}},
\label{prad}
\ee
where $d$ is the bunch spacing.

\subsection{$e^-e^-$ ideal collider \label{ideal}}

   Let us calculate the \emem luminosity, assuming an ideal superconducting collider with energy recovery, where the total available electrical power $P$ is entirely dedicated to compensating for the beamstrahlung energy losses, $P_{rad}$.

   The \emem luminosity is given by
\be
L \approx \frac{N^2c R}{4\pi\sigma_x\sigma_y d} =  k_0 \frac{N^2}{\sigma_x\sigma_z^{1/2}d}, \,\,\,\,k_0=\frac{c\gamma^{1/2}R}{4\pi\E_{ny}^{1/2}},
\label{lumee}
\ee
where the vertical beam size at the IP is
\be
\sigma_y \approx \sqrt{\epsilon_{ny} \sigma_z / \gamma} \quad (\text{assuming } \beta_y = \sigma_z),
\label{sigy}
\ee
$R \sim 0.8$ is the geometric reduction factor. By substituting the expression for $\sigma_x\sigma_z^{1/2}$ from Eq.(\ref{prad}) into Eq.(\ref{lumee}), we obtain the luminosity of an ideal \emem collider
\be
L^2=\frac{k_0^2NP}{k_3d}= \frac{3cR^2 \eta_{\text{rf}}PN}{160 \pi^2\E_{ny}r_e^3E_0d}.
\label{l-ideal}
\ee

  The beam sizes can be determined as follows. The expression for beamstrahlung losses, Eq.(\ref{prad}), yields the following combination:
\be
  \sigma_z\sigma_x^2 = k_3N^3/Pd.
\label{szsx2}
\ee
   A second combination is related to beam-beam repulsion and is determined by the vertical disruption parameter $D_y$~\cite{Chen-Yokoya}:
\be
D_y = \frac{2Nr_e\sigma_z}{\gamma\sigma_x\sigma_y}.
\label{dy}
\ee
For \emem collisions, the maximum stable value for \mbox{$D_y \approx 5$}~\cite{Thompson}. While this limit does not affect the luminosity calculated above, it does determine the maximum allowable bunch length. By combining Eqs.~(\ref{szsx2}), (\ref{dy}) and (\ref{sigy}), we obtain:
\be
\sigma_x^4=\frac{40 E_0 r_e^5 c N^5}{3 \eta_{\text{rf}}P D_y^2 \E_{ny} d},
\label{sx}
\ee
\be
\sigma_z^2=\frac{5 E_0 r_e c N D_y^2\gamma^2 \E_{ny}}{6\eta_{\text{rf}}Pd}.
\label{sz}
\ee

 Thus, for a given beamstrahlung power, we have determined the values of $L, \sigma_x,\sigma_y,\sigma_z$. The only free parameter is the bunch population $N$. According to Eq.(\ref{lumee}), the luminosity scales as $L \propto \sqrt{PN/d}$. In a practical collider, the maximum (optimal) value of $N$ is limited by the High-Order Mode (HOM) losses. As shown below, the typical optimal value is \mbox{$N\sim 10^9$}.

 As an example, for \mbox{$N=10^9$}, \mbox{$d=23$} cm (\mbox{$f_{\rm RF}=1.3$} GHz), \mbox{$\E_{ny}=3\times$$10^{-8}$} m, \mbox{$2E_0=500$} GeV, \mbox{$H_{D}=0.8$}, \mbox{$P=P_{rad}$} = 50 MW, \mbox{$\eta_{\text{rf}}=0.5$}, we obtain a luminosity $L$\,=\,3.9$\times$$10^{36}$\cms.

In this example, the average current per beam is $I$\,=\,0.2 A. The total power of the accelerated beams reaches $2P_{beam}=2\times0.2 \times 250\cdot10^9$\,=\,100 GW. The power lost to beamstrahlung at the IP $P_{rad}=P \eta_{\text{rf}}/2=50\times0.5/2=12.5$ MW. This implies an injection energy of $E_{in}$\,=\,$P\eta_{\text{rf}}/2I$\,=\,60 MeV, which is only 2.4$\times$$10^{-4}E_0$.

Consequently, such a 250 GeV collider requires an average external energy gain of only 60 MeV per electron; all subsequent energy is provided by the decelerating counter-beam. In practice, however, the energy loss spectrum is broad: while some electrons do not emit photons at all, others radiate up to several GeV. Therefore, the accelerated beam requires additional external RF power to compensate for these losses up to relatively high energies.

Below, we perform a similar calculation for the \emem ERLC collider, taking into account both cryogenic and HOM losses.

\subsection{$e^-e^-$ pulsed mode }

  Operation in pulsed mode ($DF$\,<\,1) is unavoidable when the available power is insufficient to operate the collider in continu\-ous-wave (CW) mode, even in the absence of beams. In contrast to Eq.(\ref{abd}), which accounts for RF and HOM losses, we now include a third component: beamstrahlung energy losses. Consequently, the total power in the \emem  mode consists of three primary contributions:

     \be
    P_{tot}= \left(k_1 + k_2 \frac{N^2}{d}+k_3 \frac{N^3}{\sigma_z\sigma_x^2d}\right) \times DF.
    \label{k1k2k3}
    \ee
The \emem luminosity in the pulsed mode is given by (see Eq.(\ref{lumee}))
\be
L = k_0 \frac{N^2}{\sigma_x\sigma_z^{1/2}d} \times DF.
\label{lum1}
\ee
Substituting the expression for $\sigma_z\sigma_x^2$ from  Eq.(\ref{k1k2k3}) into Eq.(\ref{lum1}) yields
\be
L^2=\frac{k_0^2N (P\times DF - (k_1+k_2N^2/d)\times (DF)^2)}{k_3d}.
\label{L2}
\ee
The maximum luminosity with respect to the duty factor $DF$  is given by
\be
L^2=\frac{k_0^2NP^2}{4(k_1d+k_2N^2)k_3}
\label{L2opt}
\ee
which is achieved (for $DF<1$) when
\be
DF=\frac{P}{2(k_1+k_2N^2/d)}.
\label{DCopt}
\ee
Substituting  Eqs.(\ref{L2opt}) and (\ref{DCopt}) into Eq.(\ref{lum1}) yields
\be
\sigma_x^2\sigma_z=\frac{N^3k_3}{k_1d+k_2N^2}.
\label{y2a}
\ee
By optimizing the luminosity with respect to $N$, we obtain:
\be
L=\frac{k_0P}{2\sqrt{2}k_3^{1/2}(k_1k_2d)^{1/4}}\,\,\,{\rm at}\,\,\, N=\sqrt{\frac{k_1d}{k_2}}.
\label{lumeedc}
\ee
Under this optimal condition, the duty factor is given by
\be
DF=\frac{P}{4k_1},
\label{DC}
\ee
and the combination of beam sizes scales as:
\be
\sigma_x^2 \sigma_z = \frac{k_1^{1/2}k_3 d^{1/2}}{2k_2^{3/2}}.
\label{y2}
\ee

By combining the expression for the disruption parameter (\ref{dy}) with Eqs. (\ref{y2a}), (\ref{y2}), and (\ref{sigy}), the values of $\sigma_x, \sigma_y$ and $\sigma_z$ can be determined independently.
Notably, under these optimal conditions, the power components are distributed such that $P_{\rm RF,dis}=P_{\rm HOM}=0.5P_{rad}=0.25P_{tot}$.

Substituting the expressions for $k_0$ and $k_3$ into Eq.(\ref{lumeedc}), we obtain the optimal (maximum) luminosity:
\be
L=0.0154\frac{(c\eta_{\text{rf}})^{1/2}R P}{\E_{ny}^{1/2}(k_1k_2d)^{1/4}E_0^{1/2}r_e^{3/2}}.
\label{lumee-dc}
\ee

  It is evident that the \emem luminosity depends very weakly on the properties of the SC linac, since $L \propto 1/(k_1k_2d)^{1/4}$, where\-as for the \epem case, the dependence is stronger: $L \propto 1/(k_1k_2d)^{1/2}$. Similar to the \epem configuration, the luminosity is independent of the accelerating gradient, and its dependence on the quality factor is even weaker: $L \propto Q_0^{1/4}$. Furthermore, Eq.(\ref{lumee-dc}) shows that in pulsed mode, the \emem luminosity scales as $L \propto P/E_0$, since  both $k_1$ and $k_2$ are proportional to the energy.

 \subsection{$e^-e^-$ CW mode}

   In this case, we follow the arguments of the previous section but set the duty factor to $DF=1$.
Substituting this into Eq.(\ref{L2}), we obtain:
\be
L^2 = \frac{k_0^2 N (P - k_1 - k_2 N^2/d)}{k_3 d}.
\label{L2cw}
\ee
By combining Eq.(\ref{L2cw}) with Eq.(\ref{lumee}), we find:
\be
\sigma_x^2 \sigma_z = \frac{N^3 k_3}{P d - k_1 d - k_2 N^2}.
\label{y2cw1}
 \ee
 The maximum luminosity with respect to the bunch population $N$ is given by
 \be
 L^4=\frac{4k_0^4(P-k_1)^3}{27k_3^2k_2d}
\label{L4}
 \ee
 This maximum is achieved when
\be
N= \sqrt{\frac{(P-k_1)d}{3k_2}}.
\ee
Under these optimal conditions, the combination of beam sizes is given by
\be
\sigma_x^2\sigma_z = \frac{k_3 (P-k_1)^{1/2}d^{1/2}}{2\sqrt{3}\,k_2^{3/2}}.
\label{y2cw2}
\ee

 As before, by combining the expression for the disruption parameter (\ref{dy}) with Eqs.(\ref{y2cw1}),(\ref{y2cw2}), and (\ref{sigy}), one can determine the values of $\sigma_x$, $\sigma_y$ and $\sigma_z$ separately.

In the \emem CW mode with an optimal bunch population $N$, the total power $P_{tot}$
is distributed among RF losses, HOM losses, and beamstrahlung losses as follows:
\be
k_1 \;\;\;:\;\;\; \frac{P-k_1}{3}\;\;\;:\;\;\; \frac{2}{3}(P-k_1).
\label{powers}
\ee
Substituting the expressions for $k_0$ and $k_3$ into Eq.(\ref{L4}), we obtain the optimal (maximum) luminosity in the \emem
 CW mode:
\be
L=0.027\frac{(c\eta_{\text{rf}})^{1/2}R (P-k_1)^{3/4}}{\E_{ny}^{1/2}(k_2d)^{1/4}E_0^{1/2}r_e^{3/2}}.
\label{lumee-cw}
\ee

Continuous operation requires a threshold power $P>k_1$, where $k_1\propto G$. At power levels significantly exceeding this threshold, the behavior becomes similar to that of the pulsed mode, where the specific luminosity $L/P$ is independent of $G$ and remains proportional to $Q_0^{1/4}$.

\begin{table}[h]
\caption{ Parameters of \epem and \emem linear colliders ERLC (Nb$_3$Sn), $2E_0$\,=\,250 GeV}
{\renewcommand{\arraystretch}{1.} \setlength{\tabcolsep}{0.3mm}
\vspace{-0.4cm}
\small
\begin{center}
\begin{tabular}{ l  l  l  l  l  l   }  \hline \\[-0.3cm]
                        & unit           &                      \epem  & \epem     &\emem    & \emem    \\ [1.1mm]
                        &                  &                  1.3 GHz& 0.65 GHz& 1.3 GHz  &  0.65 GHz \\  \hline \\[-2mm]
Energy $2E_0$           &GeV             &                      250   &  250     & 250     & 250          \\
Luminosity ${\mathcal L}_{{\rm tot}}$\, &$10^{36}$\cms &         1.1   &  2.5    & 3.8    & 7.8   \\
$P$ (wall)             &MW              &                    150     &150       & 150     & 150  \\
Accel. grad., $G$    &MV/m            &          40                & 40       &40       &40  \\
Length $L_{\rm act}/L_{\rm tot}$ &km     &                     6.3/16  & 6.3/16    & 6.3/16   & 6.3/16   \\
$N$ per bunch           &$10^{9}$       &                     0.62    & 2.75     & 0.41    & 1.53   \\
Bunch distance          & m              &                   0.23     &0.46      & 0.23     &  0.46  \\
Rep. rate, $f$          & Hz             &          $1.3\cdot 10^9$ &$0.65\cdot 10^9$&$1.3 \cdot 10^9$ &$0.65\cdot 10^9$\\
$\epsilon_{n,\,x}$/$\epsilon_{n,\,y}$    &   $10^{-6}$ m &  10/0.03  &10/0.03  & 1/0.02 &1/0.02  \\
$\beta^*_x$/$\beta_y$  at IP      & cm               &     0.82/0.03 & 16.4/0.03 &0.21/0.019 &3/0.02  \\
$\sigma_x$ at IP       &  $\mum$        &                     0.58    &2.6      &0.092     &0.35 \\
$\sigma_y$ at IP       & nm             &                   6.2       & 6.2      & 4      &4.05  \\
$\sigma_z$ at IP       & cm             &                   0.031      &0.031    &0.019    &0.02 \\
$\Delta E/E$ at IP &  $10^{-4}$   &    0.165                  & 0.161      & 4.7     & 4.3  \\
$P_{rad}$ at IP           &  MW         & 0.53                      & 1.16       & 10      & 17  \\ \hline
\end{tabular}
\end{center}
}
\label{Table1}
\end{table}

\begin{table}[h]
\caption{Parameters of \epem and \emem linear colliders ERLC (Nb$_3$Sn), $2E_0$\,=\,500 GeV}
{\renewcommand{\arraystretch}{1.} \setlength{\tabcolsep}{0.3mm}
\vspace{-0.4cm}
\small
\begin{center} 
\begin{tabular}{ l  l  l  l  l  l   }  \hline \\[-0.3cm]
                        & unit           &                      \epem  & \epem     &\emem    & \emem    \\ [1.1mm]
                        &                  &                  1.3 GHz& 0.65 GHz& 1.3 GHz  &  0.65 GHz \\  \hline \\[-2mm]
Energy $2E_0$           &GeV             &                      500   &  500     & 500     & 500          \\
Luminosity ${\mathcal L}_{{\rm tot}}$\, &$10^{36}$\cms &         1.0   &  1.9    & 3.8    & 6.9   \\
$P$ (wall)    &MW              &                    300     &300       & 300     & 300  \\
Accel. grad., $G$    &MV/m            &          40                & 40       &40       &40  \\
Length $L_{\rm act}/L_{\rm tot}$ &km     &                     12.5/30  & 12.5/30    & 12.5/30   & 12.5/30   \\
$N$ per bunch           &$10^{9}$       &                     0.81    & 2.97     & 0.41    & 1.53    \\
Bunch distance          & m              &                   0.23     &0.46      & 0.23     &  0.46  \\
Rep. rate, $f$          & Hz             &          $1.3\cdot 10^9$ &$0.65\cdot 10^9$&$1.3 \cdot 10^9$ &$0.65\cdot 10^9$\\
$\epsilon_{n,\,x}$/$\epsilon_{n,\,y}$    &   $10^{-6}$ m &  10/0.03  &10/0.03  & 1/0.02 &1/0.02  \\
$\beta^*_x$/$\beta_y$  at IP      & cm               &     4/0.09 & 55/0.09 &0.41/0.039 &6/0.04  \\
$\sigma_x$ at IP       &  $\mum$        &                     0.91    &3.4      &0.092     &0.35 \\
$\sigma_y$ at IP       & nm             &                   7.4       & 7.4      & 4      &4.05  \\
$\sigma_z$ at IP       & cm             &                   0.09      &0.09    &0.039    &0.04 \\
$\Delta E/E$ at IP &  $10^{-4}$   &    0.078                  & 0.076      & 4.5     & 4.3  \\
$P_{rad}$ at IP           &  MW         & 0.67                      & 1.18       & 19.5      & 34  \\ \hline
\end{tabular}
\end{center}
\vspace{-5mm}
}
\label{Table2}
\end{table}
\section{Examples}

   We now provide luminosity estimates using the values for $k_1$ and $k_2$ from Ref.~\cite{Telnov-erlc}.  For Nb ILC-like cavities with $f_{\rm rf}$\,=\,1.3 GHz and $Q_0$\,=\,3$\times$\,$10^{10}$, the values $k_1$\,$\approx$\,305 MW, $k_2$\,$\approx$\,(240\,${\rm MW}/$ $10^{18}$)$\times$\,23\,cm for $2E_0$\,=\,250 GeV and $G$\,=\,20 MeV/m.

   For Nb$_3$Sn cavities, the value of $k_1$ is reduced by a factor of four due to higher cryogenic efficiency at elevated operating temperatures. Assuming BCS surface resistance (which provides a baseline for scaling) the dependence is $k_1$\,$\propto$\,$f_{\rm rf}$, $k_2$\,$\propto$\,$f_{\rm rf}^2$, $d$\,$\propto$\,$1/f_{\rm rf}$. As established earlier, the general scaling with energy and gradient is  $k_1$\,$\propto$\,$E_0 G$ and $k_2$\,$\propto$\,$E_0/G$.

    Further improvements can be achieved by employing trave\-ling-wave (TW) structures \cite{Belomestnykh,Shemelin} instead of the standing-wave (SW) cavities used in the ILC. For instance, using TW cavities with an aperture radius of 3.5 cm (identical to the ILC), one can achieve a 1.3-fold increase in the accelerating gradient $G$ and a 1.7-fold reduction in RF heat losses (parameter $k_1$)~\cite{Shemelin}. Such a reduction in RF losses is critical, as it enables the continuous-wave (CW) operation of the ERLC at $2E_0$\,=\,250 and 500 GeV with high accelerating gradients and manageable power consumption.

\begin{figure}[h]
\centering
\includegraphics[width=8.5cm]{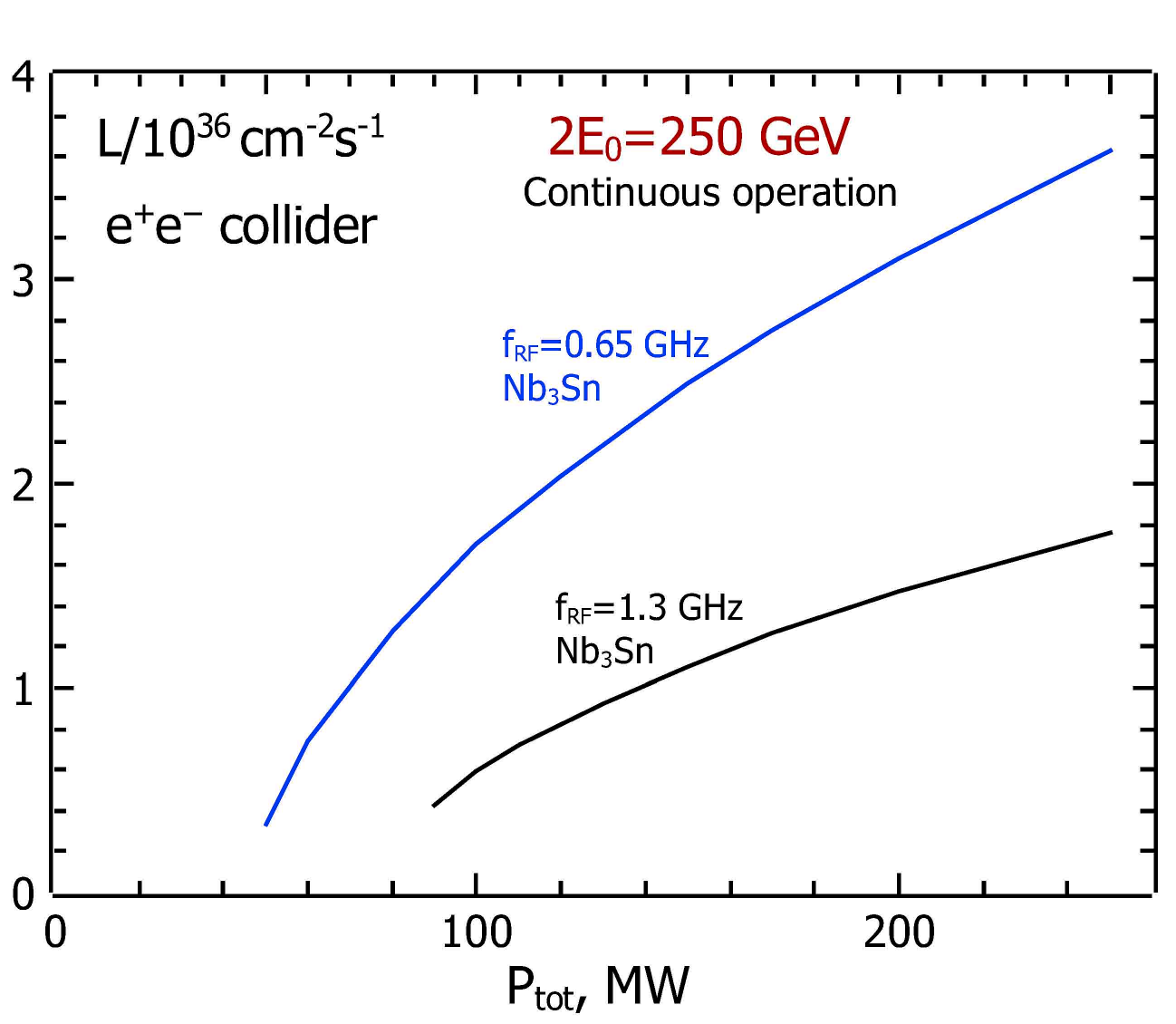}
\includegraphics[width=8.8cm]{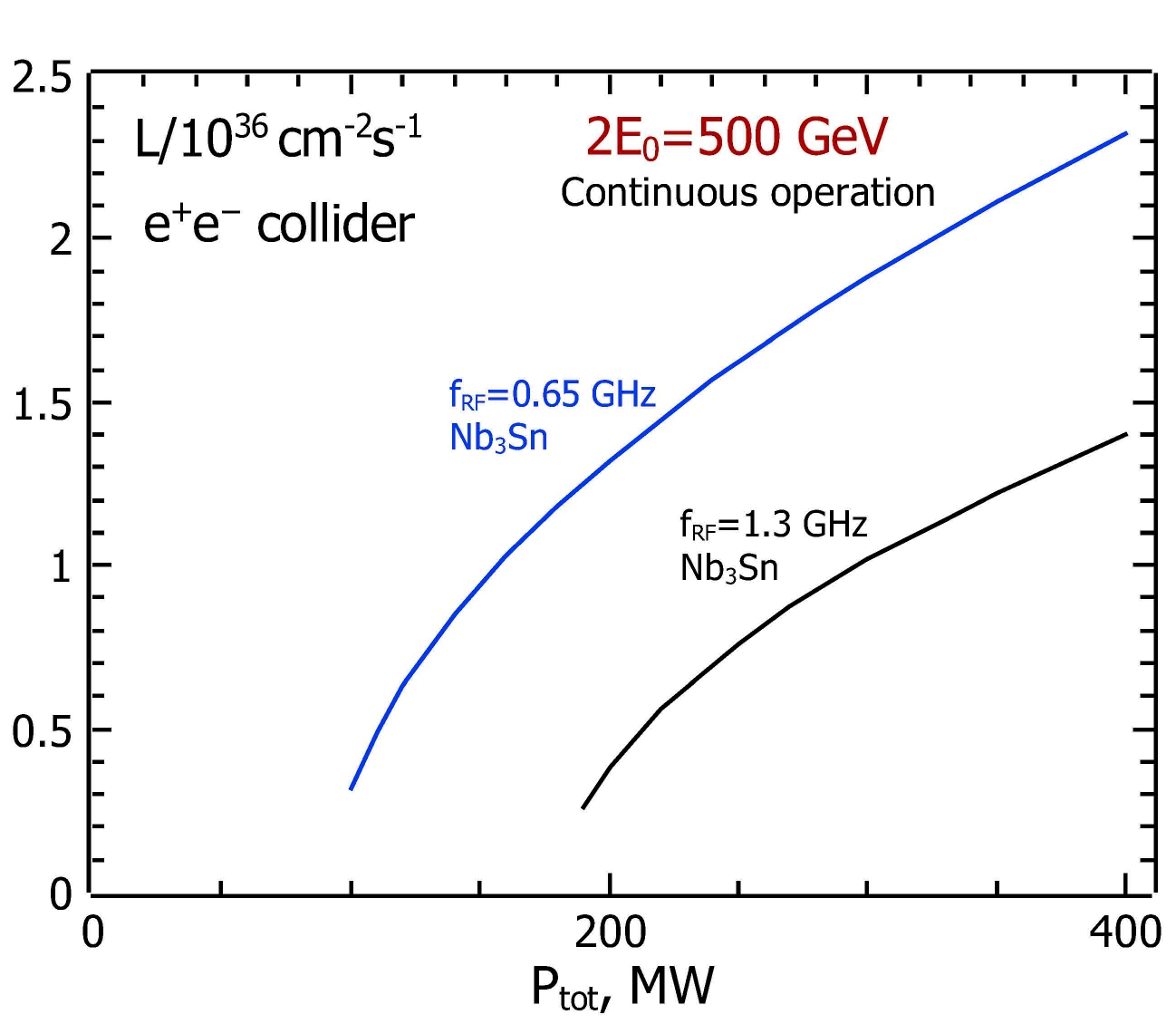}
\caption{Dependence of the $e^+e^-$ luminosity on the total power for $2E_0$\,=\,250 GeV (upper) and $2E_0$\,=\,500 GeV (bottom) in continuous mode of operation (CW), see the text.}
\label{fig-e+e-}
\end{figure}

\begin{figure}[h]
\centering
\includegraphics[width=8.5cm]{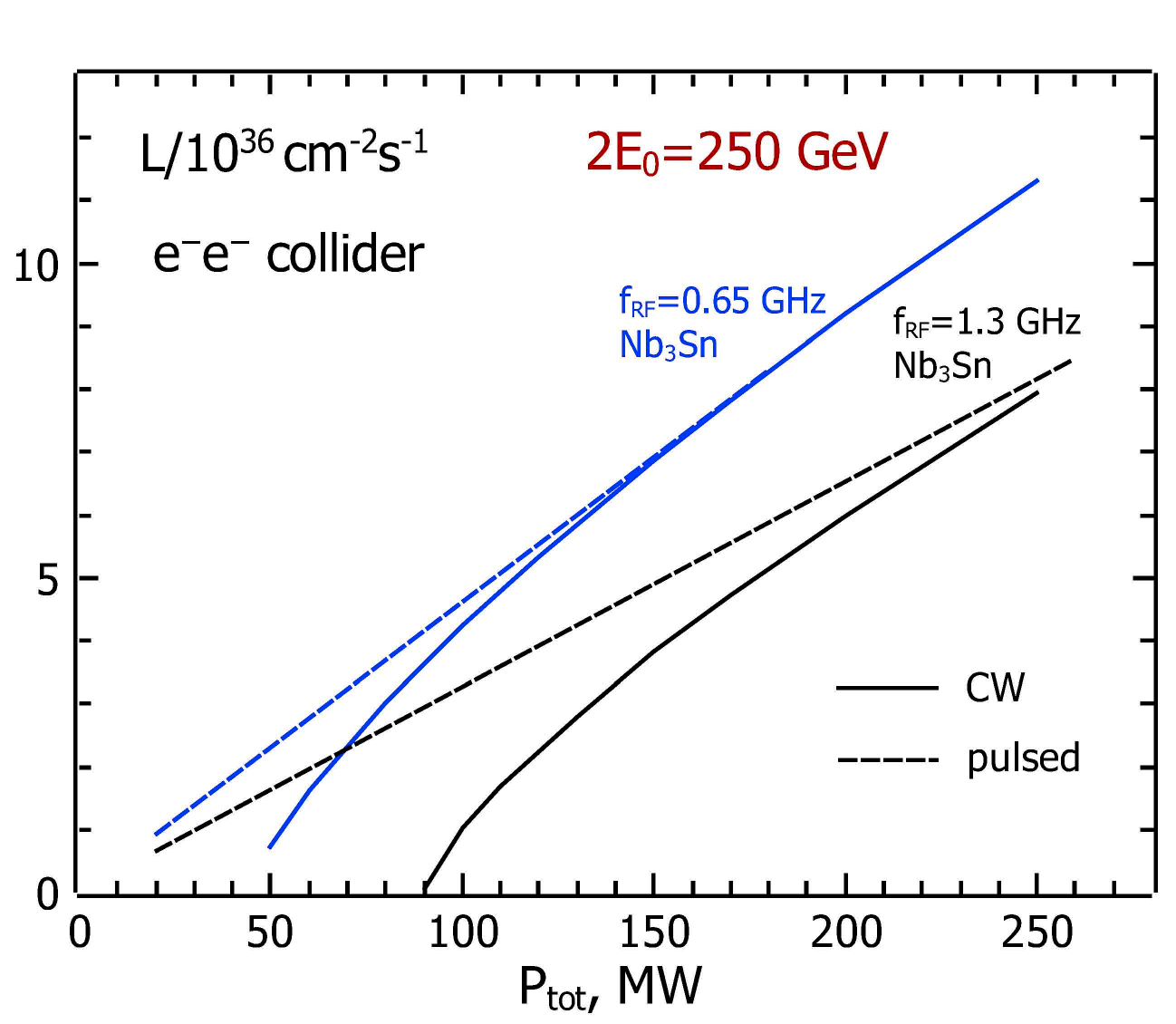}
\includegraphics[width=8.5cm]{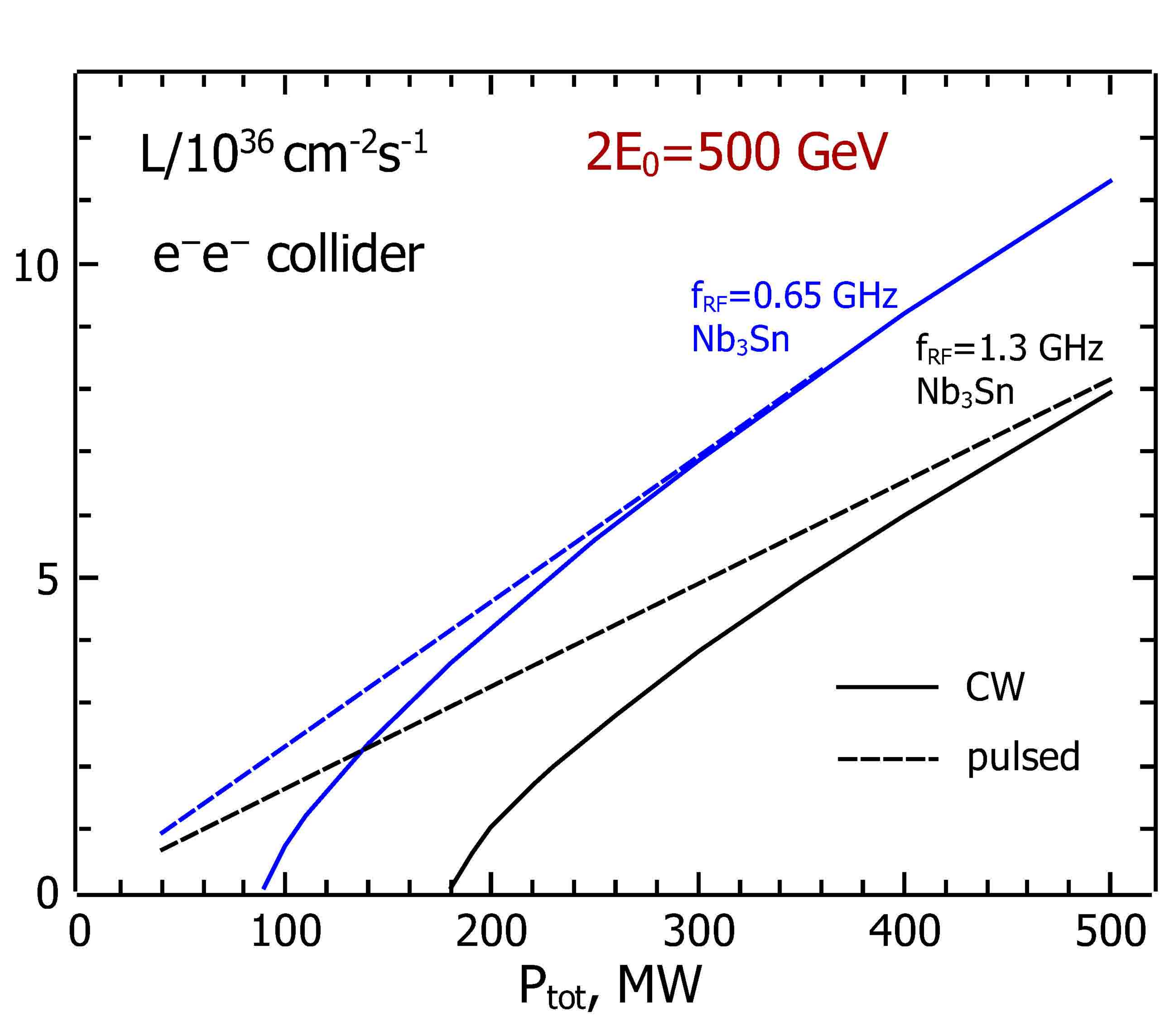}
\caption{Dependence of the $e^-e^-$ luminosity on the total power for $2E_0$\,=\,250 GeV (upper) and $2E_0$\,=\,500 GeV (bottom) in continuous(CW) and pulsed modes of operation, see the text.}
\label{fig-e-e-}
\end{figure}

The high-luminosity ERLC could be considered a candidate for a future machine, potentially utilizing the same tunnel as the ILC~\cite{Yokoya,LC-Vision}. In this scenario, the accelerating gradient should be at least as high as that of the ILC, for instance, $G$\,=\,40\,MeV/m. The most promising (though still emerging) technologies include Nb$_3$Sn ($f$\,=\,1.3 or 0.65 GHz) and traveling-wave (TW) RF structures. Operating in continuous-wave (CW) mode is advantageous as it is more compatible with cryogenic cooling systems. We assume normalized beam emittances of $\epsilon_{n,\,x}$/$\epsilon_{n,\,y}$=10/0.03 $\mu$m, similar to ILC parameters; such beams can be generated by photocathode guns. Based on these assumptions and a long-term vision for the field, we present potential parameters for \epem and \emem colliders at $2E_0=$ 250 and 500 GeV in Tables \ref{Table1},\ref{Table2} and Figures \ref{fig-e+e-},\ref{fig-e-e-}.

\section{Discussion}
    It can be seen that the achievable luminosities are $L_{\epem}$\,=\,(1--2.5)$\times$$10^{36}$ and   $L_{\emem}$=\,(3--7)$\times$$10^{36}$\cms for reasonable total powers of 150 MW and 300 MW at $2E_0$\,=\,250 and $2E_0$\,=\,500 GeV, respectively. Compared to $f_{\rm RF}$\,= 1.3 GHz, cavities operating at 0.65 GHz enable CW operation at approximately half the power and provide roughly double the luminosity, albeit at the cost of requiring wider tunnels.

The \epem collider is more complex, as the same beams must be reused many times. This necessitates measures to preserve the beam emittance over many cycles. In effect, such a machine can be viewed as a storage ring that incorporates additional components, such as linear accelerators and bunch (de)compres\-sors, within its lattice.

The \emem ERLC is significantly simpler, as each bunch is used only once and can be generated anew for each cycle. Given its potential for very high luminosity, it represents a highly attractive option. However, the \emem configuration also faces specific challenges. In the examples provided, the average injector current ranges from $I$=0.1--0.2 A, requiring high-current photocathode guns capable of producing flat beams with low emittances. These bunches must be accelerated to energies of approximately 50--100 MeV before energy recovery becomes effective (see Section \ref{ideal}). The electrical power required for the injectors is estimated at 50--100 MW (following Eq.(\ref{powers})). In our model, the beamstrahlung power at the IP is one-quarter of the injector power, amounting to 10--25 MW. For comparison, the FCC-ee design assumes 100 MW of synchrotron radiation power produced in the rings; in our case, the radiation arises solely from beam-beam collisions. This radiation power can be reduced by increasing $\sigma_x$ or $\sigma_z$ or by decreasing $N$, with the luminosity scaling as $L_{ee}$\,$\propto$\,$\sqrt{NP_{rad}}$
(see Eq.(\ref{l-ideal})).

The radiation power from the IP in each direction is approximately 10--15 MW (see Tables \ref{Table1} and \ref{Table2}). The angular spread of the photons is about $10^{-5}$, with each electron emitting roughly 0.2 photons on average. With a mean photon energy of 300--600 MeV, the total flux reaches (1--2)$\times$$10^{17}$ photons/s.

The feasibility of dumping such an intense beam must be carefully assessed. A similar problem was previously considered for photon colliders \cite{shekhtman}, and a comparable approach can be applied here. A possible scheme is as follows: at a distance of 200 m (or further) from the IP, the photon beam passes through a thin Be-Al foil and enters a 150-meter-long vessel filled with argon (Ar) at a pressure of 3--5 atm, corresponding to 4--5 radiation lengths. Due to multiple scattering in the gas, the transverse size of the electromagnetic (EM) shower becomes sufficiently wide to be dissipated in a conventional water-based ($\rm H_2O$) beam dump.

The critical component here is the input foil. At the foil's location, the photon beam is highly concentrated, with $\sigma_{x,y}$\,$\approx$\,2 mm. To manage this, the foil could be designed with a variable thickness: very thin at the center (e.g., 100 $\mum$) and increasing at larger radii where the beam density is lower. In a 100 $\mum$ Be-Al foil, photons are converted into \epem pairs with a probability of $\sim$\,0.6$\times$\,$10^{-3}$. The power deposited by ionization losses from these pairs for a photon flux of 2$\times$\,$10^{17}$ $s^{-1}$
 is approximately 1.5 W. This heat can be effectively dissipated via thermal conduction to a metal support structure at a radius of 1--1.5 cm.

      As for high-current electron sources, photoguns based on high-QE (quantum efficiency) cathodes already exist, capable of providing an average current of 0.1 A with low emittances in the range of 0.5--1 $\mu$m~\cite{Gulliford,Dunham,Rahman}. Furthermore, a cryogenic copper RF gun developed for XFEL applications can produce beams with emittances of 0.045--0.12 $\mu$m for charges of $Q$\,=\,0.1--1 nC; these beams can be made flat while preserving the emittance product~\cite{Robles}. While the performance of such a gun exceeds the requirements, its current repetition rate is too low due to limited quantum efficiency. Nevertheless, the current state of the art suggests that the required beam injector can be realized. Although beam polarization is desirable for certain physical processes, it is not strictly mandatory for the operation of the machine.

\section{Conclusion}

  Twin \epem and \emem linear colliders with energy recovery (ERLC) open the way to unprecedentedly high luminosities. In this paper, we have demonstrated that the luminosity of ERLC machines is independent of the accelerating gradient and depends only weakly on the cavity quality factor: $L_{\epem}$$\propto$\,$Q_0^{1/2}$ and $L_{\emem}$$\propto$\,$Q_0^{1/4}$.

While our previous work \cite{Telnov-erlc} focused on the \epem ERLC with bunch reuse, this paper introduces the first detailed analysis of the
\emem ERLC configuration with single-pass bunches. We show that achievable \epem luminosities reach $L_{\epem}$\,=\,(1--2.5)$\times$$10^{36}$ \cms for $2E_0$\,=\,250--500 GeV,  with \emem collisions providing approximately three times higher values. Such energy-recovery colliders represent a compelling high-lumino\-sity upgrade path for the ILC or a powerful alternative for future Higgs factories.

This proposal is currently in its early stages and requires comprehensive study and development of all core components. These include twin superconducting cavities with high $Q_0$ and high accelerating gradients operating at elevated temperatures (e.g., Nb$_3$Sn) as well as the mitigation of high-order-mode losses. Further research is needed in beam dynamics and instabilities, bunch compressor-decompressor systems, damping wigglers, and injection systems for both electrons and positrons. Additionally, the development of high-power, low-emittance electron sources and robust beam dumps for electrons, positrons, and photons is essential. The growing number of potential applications for superconducting energy-recovery accelerators~\cite{Adolphsen} suggests that rapid progress in this field is likely, with high-luminosity ERLC
colliders being the most exciting long-term goal.

  \section*{Acknowledgment}
  This work was supported by the Russian Science Foundation (grant number 24-22-00288).

\end{document}